\documentstyle[prl,aps,multicol,epsfig]{revtex}
\begin{document}
\draft

\renewcommand{\narrowtext}{\begin{multicols}{2}}
\renewcommand{\widetext}{\end{multicols}}
\newcommand{\be}{\begin{equation}}
\newcommand{\ee}{\end{equation}}
\newcommand{\lsim}   {\mathrel{\mathop{\kern 0pt \rlap
  {\raise.2ex\hbox{$<$}}}
  \lower.9ex\hbox{\kern-.190em $\sim$}}}
\newcommand{\gsim}   {\mathrel{\mathop{\kern 0pt \rlap
  {\raise.2ex\hbox{$>$}}}
  \lower.9ex\hbox{\kern-.190em $\sim$}}}
\def\be{\begin{equation}}
\def\ee{\end{equation}}
\def\ba{\begin{eqnarray}}
\def\ea{\end{eqnarray}}
\def\d{{\rm d}}
\def\i{{\rm i}}
\def\e{{\rm e}}
\def\ap{\approx}
\def\stot{\sigma_{\rm tot}}
\def\sel{\sigma_{\rm el}}
\def\skk{\sigma_{N\nu}^{\rm KK}}
\def\sSM{\sigma_{N\nu}^{\rm SM}}
\def\Ms{M_{\rm st}}
\def\Mp{M_{\rm Pl}}
\def\tr{\mbox{tr}}

\font\menge=bbold9 scaled \magstep1
\def\nota#1{\hbox{$#1\textfont1=\menge $}}
\def\R{\nota R}

\def\T{{\cal T}}

\def\d{{\rm d}}
\def\i{{\rm i}}
\def\e{{\rm e}}
\def\ap{\approx}
\def\stot{\sigma_{\rm tot}}
\def\sel{\sigma_{\rm el}}
\def\skk{\sigma_{N\nu}^{\rm KK}}
\def\sSM{\sigma_{N\nu}^{\rm SM}}
\def\Ms{M_{\rm st}}
\def\Mp{M_{\rm Pl}}
\def\Im{{\rm Im}\,}

\title{\hfill \begin{minipage}{4cm}
              {\small CERN-TH 2001-251}\\
              {\small OUTP-01-48-P}
              \end{minipage}
       \vskip1.0cm
       Remarks on the high-energy behaviour of cross-sections in
       weak-scale string theories}

\author{M.~Kachelrie{\ss}$^1$ and M.~Pl\"umacher$^2$}
\address{$^1$TH Division, CERN, CH-1211 Geneva 23, Switzerland}
\address{$^2$Theoretical Physics, University of Oxford, 
         1 Keble Road, Oxford, OX1 3NP, UK}

\date{\today}

\maketitle

\begin{abstract}
We consider the high-energy behaviour of processes involving
Kaluza-Klein (KK) gravitons of weak-scale string theories.  We discuss
how form-factors derived within string theory modify the couplings of
KK gravitons and thereby lead to an exponential fall-off of 
cross sections in the high-energy limit.
Further, we point out that the assumption of
Regge behaviour for a scattering amplitude in the high energy limit, 
$\T\propto s^{\alpha(t)}$, combined with a linear growth of the total
cross-section, $\stot(s)\propto s$, violates elastic
unitarity. Regge behaviour leads to a stringent bound on the
growth of the total 
cross-section, $\stot (s) \leq 32\pi \alpha' \ln(s/s_0)$.     
\end{abstract}


\narrowtext

\section{Introduction}
Neutrinos are the only known stable particles that can traverse
extragalactic space without attenuation even at energies $E\sim
10^{20}$~eV, thus avoiding the Greisen--Zatsepin--Kuzmin 
cutoff~\cite{general}. Since neutrinos are not deflected by (extra-)
galactic magnetic fields, this primary candidate could also explain
possible correlations between the arrival directions of observed
ultrahigh energy cosmic rays and astrophysical objects at
cosmological distances~\cite{cor}. Therefore, it has been
speculated that the ultrahigh energy primaries initiating the
observed air showers are not protons, nuclei or photons but
neutrinos~\cite{alt,do98,ja00}.  
However, in the Standard Model (SM) neutrinos are deeply
penetrating particles  producing only horizontal not vertical
extensive air showers. Consequently, either the neutrino has to be
converted locally into strongly interacting particles~\cite{We99/Fa99}
or one has to postulate new interactions that enhance the ultrahigh
energy  neutrino-nucleon cross-section.

A particular realization of the latter possibility are string theories
with $\delta$ large extra dimensions~\cite{ex}. If the
SM particles are confined to the usual $3+1$-dimensional space and
only gravity propagates in the higher-dimensional space, the
compactification radius $R$ of the extra dimensions can be
large, corresponding to a small scale $1/R$ of new physics. 
From a four-dimensional point of view the higher dimensional graviton
in these theories appears as an infinite tower of Kaluza-Klein (KK)
excitations with masses squared $m_{\vec{n}}^2=\vec{n}^2/R^2$.  Since
the weakness of the gravitational interaction is partially compensated
by the large number of KK states and cross-sections of reactions
mediated by spin 2 particles are increasing rapidly with energy, it
has been argued in Refs.~\cite{do98,ja00} that neutrinos could
initiate the observed vertical showers at the highest energies.

In Refs.~\cite{Nu99,Ka00}, the neutrino-nucleon cross-section via the
exchange of KK gravitons was calculated within the effective
field-theoretic model valid below the string scale $\Ms$. Since
amplitudes involving virtual exchange of KK gravitons diverge for more
than one large extra dimension, a form-factor suppressing the KK modes
above $\Ms$ was employed in~\cite{Ka00}. Moreover, the bound
$\stot(s)\propto\ln^2(s/s_0)$ was derived with the eikonalization
method assuming the validity of the Regge picture as an effective
theory above $\Ms$~\cite{Ka00}. 
Recently, both the use of a form-factor in the field-theoretic framework 
as well as the Froissart-like bound in the Regge picture was criticised in
Ref.~\cite{Nu01}. It is the purpose of this short article to discuss
these criticisms.  
In Sec.~II, we review how form-factors suppressing the couplings of
KK gravitons with large squared four-momentum appear in string
theory, and in Sec.~III we discuss bounds on the neutrino-nucleon
cross-section if the validity of the Regge picture above $\Ms$ is 
assumed\footnote{We do not consider the production of black holes in
neutrino-nucleon scattering in this paper \cite{blabla}.}.

\section{Form-factors for KK graviton couplings from string theory}
The amplitude for the exchange of KK gravitons in the  $t$ channel
between two particles with four momentum $p$ and $k$ is
\be \label{T}
 \T(s,t) = \frac{1}{\bar \Mp^2} \sum_{\vec n=0}^\infty 
 \bar T_{\mu\nu}(k) \,
 \frac{f^2_{\vec{n}}}
 {t-m_{\vec n}^2} \,
 \bar T^{\mu\nu}(p) \,,
\ee
where $\bar T_{\mu\nu}=T_{\mu\nu}-\frac{1}{2}\eta_{\mu\nu}
T_\alpha^{\phantom{\alpha}\alpha}$, $T_{\mu\nu}$ is their energy-momentum
tensor, $\eta_{\mu\nu}$ is the metric tensor, $\bar\Mp^2=G_N^{-1}$ is
the reduced Planck mass and $s,t$ are the usual
Mandelstam variables.

In the calculation of Ref.~\cite{Ka00}, an exponential suppression of 
the effective coupling of the level $\vec{n}$ KK mode to
four-dimensional fields was used as form-factor $f_{\vec{n}}$,
\begin{equation}    \label{f}
  f_{\vec{n}}=
  \exp\left(-{c\,m_{\vec{n}}^2\over 2\Ms^2}\right)\,,
\end{equation}
where $c$ is a constant of order $1$. Then the summation over $n$
which for $f_{\vec n}=1$ only 
converges in the case of one extra dimension becomes well-defined for
all $\delta$. However, a legitimate question to ask is whether this 
modification 
is well-motivated. To address this point, we first note that the
suppression of the couplings sets in only for $-t\gsim\Ms^2$. Thus, the
form-factor~(\ref{f}) does not modify the effective theory in its range
of validity, $-t\lsim\Ms^2$. On the other hand, 
string theories generically predict an exponential suppression of the
coupling to higher KK modes in the regime $-t\gg\Ms^2$~\cite{twist}.  
This suppression is also consistent with the idea that recoil
effects due to the finite tension of D-branes become important in the
emission of KK gravitons with large momentum transfer,
$-t\gg\Ms^2$~\cite{fluct}. 

Let us now discuss in detail how form-factors to the coupling of KK
gravitons appear as string corrections to the well-known
field theoretical result~\cite{field}. The effective low-energy coupling
of standard model particles to KK gravitons can be obtained by
computing the amplitude for the scattering of four open strings. 
At tree-level the string amplitude is given by
\begin{eqnarray}  \label{tree}
 A_{\rm tree}&=&
   g^2A(1,2,3,4)\;\mbox{tr}[t^1t^2t^3t^4+t^4t^3t^2t^1]\;S(s,t)\nonumber\\[1ex]
&+& g^2A(1,3,2,4)\;\mbox{tr}[t^1t^3t^2t^4+t^4t^2t^3t^1]\;S(s,u)\nonumber\\[1ex]
&+& g^2A(1,2,4,3)\;\mbox{tr}[t^1t^2t^4t^3+t^3t^4t^2t^1]\;S(t,u)\;,
\end{eqnarray}
where the $t^i$ are Chan-Paton matrices, $g^2A(i,j,k,l)$ are the amplitudes
evaluated in the low-energy field theory framework, and $S(s,t)$ is
the Veneziano amplitude,
\begin{equation} \label{veneziano}
S(s,t)={\Gamma\left(1-{s\over\Ms^2}\right)
        \Gamma\left(1-{t\over\Ms^2}\right) \over
        \Gamma\left(1-{s\over\Ms^2}-{t\over\Ms^2}\right)}\;.
\end{equation}
The Veneziano amplitude has the usual Regge-pole structure. In the 
hard-scattering limit, $s\to\infty$ with fixed scattering angle
$t/s=-\sin^2{\phi\over2}$, the amplitude falls off exponentially,
\begin{equation} \label{exp}
  S(s,t)\propto \mbox{exp}\left[ {s\over \Ms^2} \left(
        \mbox{sin}^2{\phi\over2}\,\mbox{ln}\,\mbox{sin}^2{\phi\over2}
       +\mbox{cos}^2{\phi\over2}\,\mbox{ln}\,\mbox{cos}^2{\phi\over2}
       \right)\right]  \,.
\end{equation}
It was pointed out in Ref.~\cite{hasch} that this exponential fall-off
can be interpreted as an effective thickness of D-branes of the order
of the string scale, which would give rise to a form-factor of the kind
(\ref{f}).

However, this tree-level result is due to string Regge excitations.
Contributions from KK gravitons first appear at the one-loop level
\cite{love,nonplanar,schwarz}. 
The Type I string diagram which contains $s$-channel gravitational
exchange in the low-energy limit is the non-planar cylinder diagram,
which gives rise to contributions to the amplitude of the form
\begin{equation}
A_{\rm 1-loop} = -{1\over2}s\,tA(1,2,3,4)\,\tr(t^1t^2)\tr(t^3t^4)\, f^{(1)}_{\rm T}(s,t)
+ \ldots ,
\end{equation}
where the dots indicate permutations of the indices $1,\ldots,4$ and the 
Mandelstam variables $s$, $t$ and $u$.

The non-planar cylinder amplitude $f^{(1)}_{\rm T}(s,t)$ has recently been 
analysed in detail \cite{dudas,peskin}. It is given by
\begin{eqnarray} \label{f1}
f^{(1)}_{\rm T}(s,t)&=&{g_s^2\over{\Ms^{10}}}\int\limits_0^\infty\mbox{d}l
\int\limits_0^1\mbox{d}\nu_2\int\limits_0^{\nu_2}\mbox{d}\nu_1 
\int\limits_0^1\mbox{d}\nu_3 \\[1ex]
&&\times\left({\tilde{\psi}_{13}^{\rm T}\tilde{\psi}_{24}^{\rm T}
\over \tilde{\psi}_{12}\tilde{\psi}_{34}}\right)^{s/\Ms^2}
\left({\tilde{\psi}_{13}^{\rm T}\tilde{\psi}_{24}^{\rm T}
\over \tilde{\psi}_{14}^{\rm T}\tilde{\psi}_{23}^{\rm T}}\right)^{t/\Ms^2}
F_6(l,R)\nonumber\;,
\end{eqnarray}
where
\begin{equation}
F_6(l,R)={\Ms^6\over(R \Ms)^6}\vartheta_3^6\left(0,{1\over2}il(R\Ms)^2\right)
\end{equation}
are the winding mode contributions from the toroidal compactification.
For simplicity we have assumed that all six extra dimensions have
the same compactification radius $R$. Further,
\begin{eqnarray}
\tilde{\psi}_{ij}&=&{1\over l}
             {\vartheta_1(\nu_j-\nu_i,il)\over \eta^3(il)}\;,\\[1ex]
\tilde{\psi}_{ij}^{\rm T}&=&{1\over l}
             {\vartheta_4(\nu_j-\nu_i,il)\over \eta^3(il)}\;,
\end{eqnarray}
where $\vartheta_i$ are the usual Jacobi $\vartheta$ functions and $\eta$
is the Dedekind function.

In Eq.~(\ref{f1}), $l$ is the modulus of the cylinder and in the low-energy 
limit the contribution from exchange of KK gravitons in the $s$-channel can 
be extracted by considering the $l\to\infty$ limit of the integrand in 
Eq.~(\ref{f1}). In this limit one finds \cite{dudas}:
\begin{eqnarray} \label{f1_2}
&&f^{(1)}_{\rm T}(s,t)\approx
\left[{2^{-s/\Ms^2}\over\sqrt{\pi}M_{\rm Pl}}
{\Gamma\left({1\over2}-{s\over2\Ms^2}\right)\over
\Gamma\left(1-{s\over2\Ms^2}\right)}\right]^2 \\[1ex]
&&\times{1\over\Ms^2}\int\limits_0^{\infty}\mbox{d}l\sum\limits_{\vec{n}}
\mbox{exp}\left[-\pi l\left(-{s\over2\Ms^2}+{\vec{n}^2\over2}(R\Ms)^2
\right)\right]\;.\nonumber
\end{eqnarray}
The integral in the second line is just the proper-time parametrization
of the sum over the winding mode propagators. The $l\to0$ limit of the
integration is ill-defined, since we had replaced the integrand in 
Eq.~(\ref{f1}) by its asymptotic expansion for $l\to\infty$.
However, this divergence corresponds only to a harmless IR divergence 
of a box diagram~\cite{dudas}. By the modular transformation
$l={1\over\tau}$ the amplitude~(\ref{f1}) can be represented as a sum
of box diagrams giving a well-defined result in the limit $l\to0$.

Here we are mainly interested in the KK graviton contributions. 
The prefactor in Eq.~(\ref{f1_2}) can be interpreted as the squared 
form factor for the emission of a KK graviton with momentum-squared 
$s>0$, 
\begin{equation} \label{fn_2}
f_{\vec{n}}={2^{-s/\Ms^2}\over\sqrt{\pi}M_{\rm Pl}}
{\Gamma\left({1\over2}-{s\over2\Ms^2}\right)\over
\Gamma\left(1-{s\over2\Ms^2}\right)}\;.
\end{equation}
In the hard scattering limit this gives an
exponential suppression of the KK graviton coupling
\begin{equation}
f_{\vec{n}}\propto \mbox{exp}\left(-{s\over\Ms^2}{\rm ln}\,2\right)
\quad\mbox{for}\quad s\gg\Ms^2\;.
\end{equation}
This form-factor is valid for KK gravitons emitted either on-shell or for
virtual graviton exchange in the $s$-channel. In these cases we
recover Eq.~(\ref{f}) with $s=m_{\vec{n}}^2$.

Next, we discuss the $t$-channel exchange of KK gravitons which is
important for neutrino-nucleon scattering. Contributions from 
the $t$-channel exchange  are contained in the planar
cylinder diagram, which is a one-loop correction to 
the tree-level amplitude (\ref{tree}) of the form
\begin{equation}
A_{\rm 1-loop}=-{1\over2}s\,tA(1,2,3,4)\,\tr(t^1t^2t^3t^4)\, f^{(1)}(s,t) + \ldots
\end{equation}
The planar cylinder amplitude $f^{(1)}(s,t)$ is given by \cite{schwarz,planar}
\begin{eqnarray} \label{f1_p}
f^{(1)}(s,t)&=&{g_s^2\over{\Ms^{10}}}\int\limits_0^\infty\mbox{d}l
\int\limits_0^1\mbox{d}\nu_2\int\limits_0^{\nu_2}\mbox{d}\nu_1 
\int\limits_0^1\mbox{d}\nu_3 \\[1ex]
&&\times\left({\tilde{\psi}_{13}\tilde{\psi}_{24}
\over \tilde{\psi}_{12}\tilde{\psi}_{34}}\right)^{s/\Ms^2}
\left({\tilde{\psi}_{13}\tilde{\psi}_{24}
\over \tilde{\psi}_{14}\tilde{\psi}_{23}}\right)^{t/\Ms^2}
F_6(l,R)\nonumber\;.
\end{eqnarray}
Again considering the $l\to\infty$ limit of the integrand one obtains
a contribution to the amplitude which can be written as a derivative
of the tree-level Veneziano amplitude \cite{schwarz},
\begin{eqnarray} \label{f1_p2}
&&f^{(1)}(s,t)\approx {\Ms^5\over4\pi^2stM_{\rm Pl}^2}
\left({\partial S(s,t)\over\partial\Ms}\right)\\[1ex]
&&\times{1\over\Ms^2}\int\limits_0^{\infty}\mbox{d}l\sum\limits_{\vec{n}}
\mbox{exp}\left[-\pi l{\vec{n}^2\over2}(R\Ms)^2\right]\;.\nonumber
\end{eqnarray}
It is easy to see that the prefactor in Eq.~(\ref{f1_p2}) has the 
same Regge-pole structure as the Veneziano amplitude. In the 
hard-scattering limit, $s\to\infty$, $t/s$ fixed, the amplitude
again falls off exponentially, as in Eq.~(\ref{exp}).
Thus, form-factors derived within string theory indeed modify the
couplings of virtual KK gravitons and, hence, also the
cross-sections in the high-energy limit.

\section{Unitarity limits in the Regge picture}
The authors of Ref.~\cite{Nu99} pointed out that the Regge picture is a
reasonable approximation to string theory valid above $\Ms$. As
motivation for this assumption we note that the Regge picture takes
into account not only the KK modes of the graviton but also those from
lower lying trajectories and misses only genuine string modes like
winding modes.  In the following, we will therefore also use this
assumption and derive bounds on the total cross-sections valid within
this framework.

A general Regge amplitude $\T_R$ can be represented by
\be  \label{regge}
 \T_R(s,t)= \beta(t) \left(\frac{s}{s_0}\right)^{\alpha(t)} \,,
\ee
where the exponent $\alpha(t)$ is given by the Chew-Frautschi plot of
the spin against the  mass of the particles lying on the leading Regge
trajectory contributing to the reaction. In our case, the intercept
$\alpha(0)$ of this trajectory is equal to the spin $j$ of the massless
graviton, $\alpha(0)=2$.  

We first note that a Regge amplitude with intercept
$\alpha(0)=2$ gives via the optical theorem a total cross-section  
growing linearly with $s$,
\be   \label{opt}
\sigma_{\rm tot}(s) = \frac{1}{s}\,\Im\{\T_R(s,0)\}\propto s^{\alpha(0)-1}  \,.
\ee 
Thus the assumed Regge-behaviour alone, without any unitarization,
reduces the growth of 
the total cross-section by one power of $s$ compared to the naive
expectation $\sigma\propto s^j=s^2$. On the other hand, the
elastic cross-section 
\be
\sel(s) = \frac{1}{16\pi s^2} \int_{-s}^0 \d t \: |\T_R(s,t)|^2
        \propto \frac{s^2}{\ln(s/s_0)} 
\ee
increases even faster than the total cross-section. 
Therefore, elastic unitarity, $\stot\geq\sel$, is violated above a
certain energy for any Regge amplitude with $\alpha(0)>1$ ---
as is well-known from the case of the pomeron.
These findings are in clear contradiction to the ones of
Ref.~\cite{Nu99,Nu01}: there, it was claimed that a linear growth of
$\sigma_{\rm tot}(s)$ for $\T_R(s,0)\propto s^2$ respects unitarity.

Next, we derive the maximal total cross-section allowed for an
arbitrary Regge amplitude by elastic unitarity. Following
Leader~\cite{le63}, we rewrite $\T_R(s,t)$ as 
\be
 \T_R(s,t) = \beta \left(\frac{s}{s_0}\right)^{\alpha(t)} 
          = \T_R(s,0) \left(\frac{s}{s_0}\right)^{\alpha(t)-\alpha(0)}
\ee
and expand the amplitude around $t=0$,
\be
 \left(\frac{s}{s_0}\right)^{\alpha(t)-\alpha(0)} =
 \exp\{ \alpha' t \ln(s/s_0) + O(\alpha^{\prime\prime}t^2) \} \,.
\ee
Here, $\alpha'$ denotes the derivative of $\alpha(t)$ evaluated at
$t=0$ and we have neglected for clarity possible non-linear terms in
$t$ and the subdominant $t$ dependence of $\beta$. 
Then we evaluate $\sel$,
\be
 \sel = \frac{1}{16\pi s^2} \int_{-s}^0 \d t \: |\T_R(s,t)|^2
 = \frac{|\T_R(s,0)|^2}{16\pi s^2} \:\frac{1}{2\alpha' \ln(s/s_0)} \,.
\ee
Requiring now elastic unitarity 
\be
\sel \leq \stot=\frac{1}{s}\: \Im\{ \T_R(s,0) \} <\frac{1}{s}\: |\T_R(s,0)| \,,
\ee
it follows
\be
\frac{1}{32\pi\alpha' \ln(s/s_0)} \:\frac{|\T_R(s,0)|^2}{s^2} \leq  \stot 
\ee
or
\be
\frac{\stot^2}{32\pi\alpha' \ln(s/s_0)} \leq \stot 
\ee
and finally
\be   \label{bound}
\stot (s) \leq 32\pi \alpha' \ln(s/s_0) \,.
\ee
Thus the assumption of a Regge amplitude results in a stronger bound
for the total cross-section than the Froissart bound. A more general
derivation of such a bound can be found in Ref.~\cite{ed67}.

Some remarks are now in order: 
First, we have always used formulae valid for $d=4$ dimensions.
This is appropriate because the main contribution to the cross-sections
comes from the small $t$ region and therefore does not probe the
extra dimensions.  
Second, we note that this bound applies on the parton not the
hadron level.  
Third, the bound~(\ref{bound}) contains two parameters, the slope of
the Regge trajectory $\alpha'$ and the unknown scale $s_0$, and is
therefore still not useful for a {\em numerical\/} evaluation.

To proceed, we use that
\be \label{b}
 \sigma_{\rm tot}^{N\nu} (s) = 
 \left[ N(s)+ \delta N \right] \sigma_{\rm tot} (s) 
 \leq N(s)\sigma_{\rm tot} (s) \,,
\ee
where $N(s)\propto s^{0.4}$ \cite{ga95} takes into account the
increasing number of target partons in the nucleon. The term
$\delta N<0$ corrects that each parton carries only a fraction
$x<1$ of the nucleon momentum, i.e. that $\ln(xs/s_0)<\ln(s/s_0)$.
A numerical value for the bound~(\ref{b}) can now be determined 
by joining the field-theoretic result  
and the Regge result on the hadron level at that scale $s'\sim \Ms^2$, 
where the field-theoretic result starts to violate s-wave unitarity on
the parton level. We find that the  KK contribution to the total
cross-section at UHE is at most of the same order of magnitude as the 
SM cross-section.

Finally, we want to comment briefly on the suggestion of Ref.~\cite{do98}
that the exponential increase of (lepto-quark like) KK resonances in the
s channel could enhance the neutrino-nucleon cross-section. Since the
Horn-Schmid duality~\cite{HS} connects $s$ and $t$ channel Regge/String
amplitudes, our discussion above can be applied immediately to this
case. The $n=0$ lepto-quarks can have either spin $j=0$ or 1. Even in the
later case, the intercept will be smaller than 1 and the partonic
cross-section will be asymptotically decreasing with $s$.

\section{Conclusions}
Couplings of KK states derived within a field-theoretic model
valid below $\Ms$ are modified by form-factors calculable in string
theory~\cite{hasch,dudas}. The use of these form-factors makes the
sums over KK states well-defined without simply cutting-off KK modes
with $m_n\gsim\Ms$. For the case of neutrino-nucleon scattering, the 
tree-level string cross-section was calculated in Ref.~\cite{cornet}:
even for a string scale as low as 1~TeV, the cross-section found there
is only of the same order as the SM cross-section at energies
$s\gsim\Ms^2$. Taking KK graviton exchange into account as one-loop
correction yields a cross-section that is still very different
from the nucleon-nucleon cross-section, i.e.\
neutrino-nucleon scattering cannot explain the observed
ultrahigh energy cosmic rays.

We have addressed the question how the assumption of Regge behaviour,
$\T\propto s^{\alpha(t)}$, for the neutrino-nucleon scattering
amplitude above $\Ms$ bounds the growth of $\stot$. We have shown that
a linear growth of $\stot$, as advocated in~\cite{Nu99,Nu01}, is
incompatible with elastic unitarity. Regge behaviour allows instead
only logarithmic growth of the partonic cross-sections.

\acknowledgments
We are grateful to E.~Dudas, A.~Faraggi, J.~Mourad and D.~Ross for 
useful comments. M.P.\ was supported
by the EU network ``Supersymmetry and the early universe''
under contract no.~HPRN-CT-2000-00152.

\widetext

\end{document}